\documentclass[prb,11pt]{revtex4-1}

\usepackage{amsmath}    
\usepackage{graphicx}   
\usepackage{verbatim}   
\usepackage{color}      
\usepackage{subfigure}  
\usepackage{hyperref}   
\begin{document}

\title{Checking the validity of truncating the cumulant hierarchy description of a small system}
\author{Manuel Morillo, Jos\'e G\'omez-Ord\'o\~nez and Jos\'e M. Casado}
\affiliation{Universidad de Sevilla. Facultad de F\'{\i}sica. \'Area de
F\'{\i}sica Te\'orica. Apartado de Correos 1065. Sevilla 41080. Spain }

\begin{abstract}
We analyze the behavior of the first few cumulant in an array with a small number of coupled identical particles. 
Desai and Zwanzig (J. Stat. Phys., {\bf 19}, 1 (1978), p. 1) studied noisy arrays of nonlinear  units with global coupling  and derived an infinite hierarchy of differential equations for the cumulant moments. They focused on the behavior of infinite size systems using a strategy based on truncating the hierarchy. In this work we explore the reliability of such an approach to describe systems with a small number of elements.
We carry out an extensive numerical analysis of the truncated hierarchy as well as  numerical simulations of the full set of Langevin equations governing the dynamics. We find that the results provided by the truncated hierarchy for finite systems  are at variance with those of the Langevin  simulations for large regions of parameter space. The truncation of the hierarchy leads to a dependence on initial conditions and to the coexistence of states which are not consistent with the theoretical expectations based on the multidimensional linear Fokker-Planck  equation for finite arrays.
\end{abstract}

\maketitle

\section{Introduction}
\label{sec:1}
The description of nonlinear stochastic systems can hardly be carried out without approximations due to the 
interplay of noise  and nonlinearity. In some problems, the stationary distribution for the relevant variables is available in analytical form, but in general very little information can be obtained without
approximations. A convenient way of describing the system dynamics is in term of cumulant moments satisfying an infinite set of coupled ordinary differential equations \cite{RISKEN1984}. For all practical purposes, this infinite hierarchy  needs to be truncated in order to obtain a finite set of closed equations. A very much used approximation consists in truncating the infinite hierarchy at the Gaussian level by neglecting cumulants of order three and higher. For a stochastic variable $Y(t)$, Marcinkiewicz \cite{mar} indicated that the characteristic function $\Gamma(\lambda) = \langle \exp(i \lambda Y(t)) \rangle$ can be expressed as $\Gamma(\lambda) = \exp (P(\lambda,t))$ where $ P(\lambda,t)$ is a polynomial of first or second degree in $\lambda$. Consequently, as pointed out by H\"anggi and Talkner \cite{hantal}, the truncation of the hierarchy at levels higher than two is problematic, although, as these authors emphasize, it is not an empty concept. In their own words "it only means that the neglect of cumulants beyond a given order cannot be justified a priori".

In this work we consider the statistical mechanical description of a 
stochastic array containing a small finite number $N$ of coupled identical elements. The dynamics will be given by a set of $N$ coupled nonlinear stochastic differential equations for the degrees of freedom characterizing the elements of the array. Equivalently, we can describe 
the system in terms of an $N$-dimensional joint probability distribution. 
We will
assume that this joint probability distribution satisfies an $N$-dimensional linear Fokker-Planck equation (FPE).  

Obtaining dynamical information from the FPE is plagued with difficulties due
to the nonlinear character of the dynamics. As mentioned above, a way of dealing with them is to consider the cumulant moments satisfying an 
infinite hierarchy of ordinary differential equations. In the case of the arrays studied in this work, Desai and Zwanzig \cite{DZ1978} 
derived from the FPE such an infinite hierarchy. Two types of cumulants appear: diagonal cumulant
moments associated to a single degree of freedom, $M_i(t)$, and cross
cumulant moments, $\mu_{ijk\ldots}(t) $ involving two or more 
variables. For an infinite system, the off-diagonal cumulant moments are negligible 
if they were zero at the initial preparation (see \cite{DZ1978}). In other words, off-diagonal
cumulant moments are not generated by dynamical evolution in the case of infinitely large systems. But even then,  
for all practical purposes, the hierarchy has to be approximated by truncating it at a certain level. The results 
of truncating at different levels the hierarchy of diagonal cumulant moments in the $N \rightarrow \infty$ limit 
was already discussed by Desai and Zwanzig. As also indicated in \cite{hantal}, the cumulant truncation is a succesful strategy in that limit.

By contrast with the previous work of Desai and Zwanzig, the present work focuses on systems with a small number $N$ of elements. In this case, both the diagonal and off-diagonal cumulant moments have to be taken into account. It is then an open question whether
truncating the infinite hierarchy is an adequate approximation strategy.
It is not possible to solve analytically the truncated set of equations at any level. So we will rely on a numerical treatment of the set.
Our goal is to elucidate whether truncation of the infinite hierarchy of cumulant moments at different levels provides a reliable approximation for small finite arrays. To this end, we will compare the results for the first few cumulant moments obtained from the truncated hierarchy with those obtained by numerically solving the set of the $N$ coupled Langevin equations. Besides the numerical simulations of the Langevin dynamics, the 
results of the truncated hierarchy will also be contrasted with the exact results expected from the  stationary solution of the $N$-dimensional Fokker-Planck equation, as well as the global stability H-theorem. We will see that the simulation results agree with the exact known information, while the truncation strategy might lead to results incompatible with the exact results.  

\section{The model}
\label{sec:2} 
We consider a global coupling model that can be viewed as a set of nonlinear ``oscillators", each
of them described by a ``coordinate" $x_i$. The dynamics
of the system is given by the set of coupled Langevin equations (in
dimensionless form)
 \begin{equation}
 \dot{x}_{i}=x_{i}-x_{i}^3+\frac{\theta}{N}\sum_{j=1}^{N}(x_{j}-x_{i})+\xi_{i}(t),
 \label{EQ1} 
 \end{equation}
where $\theta$ is the strength of the global coupling term. This
parameter will be taken to be either positive or negative. The terms
$\xi_{i}(t),\; i=1,2,\ldots N$ represent $N$ uncorrelated
Gaussian white noises with zero averages and
$\langle\xi_{i}(t_1)\xi_{j}(t_2)\rangle=2D\delta_{ij}\delta(t_1-t_2)$. This model was introduced by Kometani and Shimizu \cite{KomShi1975} as a model for muscle contractions,
and it was later on analyzed by Desai and Zwanzig \cite{DZ1978} from a Statistical Mechanics perspective. The model
describes $N$ degrees of freedom each of them globally coupled to all the other ones. Each degree of freedom has an
intrinsic nonlinear dynamics. The nonlinearity and the presence of the noise
terms render the behavior of the system far from trivial. 

An alternative formulation of the dynamics is in terms of the
linear Fokker-Planck equation for the joint probability density
$f_{N}(x_{1},x_{2},\ldots,x_{N},t)$,
\begin{equation}
\frac{\partial f_{N}}{\partial t}=\sum_{i}\frac{\partial }{\partial
  x_i}\Big(\frac{\partial U}{\partial
  x_i}f_{N}\Big)+D\sum_{i}\frac{\partial^2 f_{N}}{\partial x_{i}^2},
\label{EQ3} 
\end{equation} 
where $U$ is the potential energy relief,
\begin{equation}
U=\sum_{i=1}^{N}V(x_{i})+\frac{\theta}{4N}\sum_{i=1}^{N}\sum_{j=1}^{N}(x_{j}-x_{i})^2,
\label{EQ4} 
\end{equation}
with the single particle potential
\begin{equation}
V(x_i)=\frac{x_i^4}{4}-\frac{x_i^2}{2}.
\label{EQ6} 
\end{equation}
The term $V(x_i)$ describes a symmetrical potential with two wells of equal depths separated by a hump at $x_i=0$. The interaction part of the full potential modifies it in such a way that for $\theta > 1$ the two wells blend into a single minimum at $x_i=0$. For $\theta <1$, the two wells exist, but their locations and the barrier height depend on $\theta$. Note that for $\theta >0$, the interaction energy contribution to the full potential favors that any pair $x_i$ and $x_j$ should have the same sign (both either positive or negative), while for $\theta <0$, the opposite happens and the interaction tends to favor configurations with positive and negative values of the variable.

The only explicit solution of Eq.~(\ref{EQ3}) is its long time stationary one, given by 
\begin{equation}
\label{LFPEQ}
f_{N}^{st}(x_{1},x_{2},\ldots,x_{N}) = \frac 1Z\, \exp \Big (-\frac U D \Big ) 
\end{equation}
where $Z$ is a normalization function. This solution is independent of the initial condition, and the system will necessarily relax to it, although the time it takes to do it might be, depending on the system parameters, extremely large. It should be noted that we keep the finite number of particles $N$ fixed, while we take the long time limit. Had we have taken the limit $N \rightarrow \infty $ first, as done in the infinite size limit studies, and afterwards the long time limit, the result would differ.

As mentioned above, Desai and Zwanzig \cite{DZ1978} 
derived an infinite hierarchy of ordinary differential equations for the
cumulant moments. Neglecting the cumulant moments of order five and higher one gets a set of 11 equations for the first four order (diagonal and off-diagonal)
cumulant moments (see the Appendix in \cite{DZ1978}). For asymptotically large systems ($ N \rightarrow \infty$), Desai and Zwanzig found that, for some regions of parameter space, zero is the only stationary value for the first cumulant moment. For other regions of parameter values, there are two stable nonzero stationary values, while the zero value becomes unstable. The regions are separated by a transition line whose shape depends on whether $\theta >1$ or $0<\theta < 1$. As we will see in the next section, the truncated hierarchy of cumulants for small systems might yield several stable stationary moments depending upon the initial preparation, as obtained within the $ N \rightarrow \infty$ limit. Nonetheless, those results are incompatible with the unicity of the solution of a multidimensional linear FPE and with the independence from initial condition of the long-time results, required by the H-theorem. 

\section{Numerical simulations}
We have carried out numerical simulations of the whole set of Langevin equations, Eq.\ (\ref{EQ1}). Using the procedure detailed in Ref. \cite{CDGMH2003}, we have integrated the Langevin equations for a
large number of noise realizations (typically 5000
realizations). Averaging over them, we estimate the first two
cumulant moments of a single variable by
\begin{equation}
M_1(t) = \langle x(t) \rangle \approx \frac 1{\mathcal {N}} \sum_{\alpha=1}^\mathcal{N} x^{(\alpha)}(t),
\end{equation}
and
\begin{equation}
M_2(t) = \langle x^2(t) \rangle -  M_1^2(t) \approx \frac 1{\mathcal {N}} \sum_{\alpha=1}^\mathcal{N} (x^{(\alpha)})^2(t)-M_1^2(t),
\end{equation}
where $\mathcal {N}$ indicates the total number of trajectories and $x^{(\alpha )}(t)$ indicates the numerically obtained single particle trajectory in the $\alpha$ noise realization.
 
\begin{figure}
\center
\includegraphics[width=9cm]{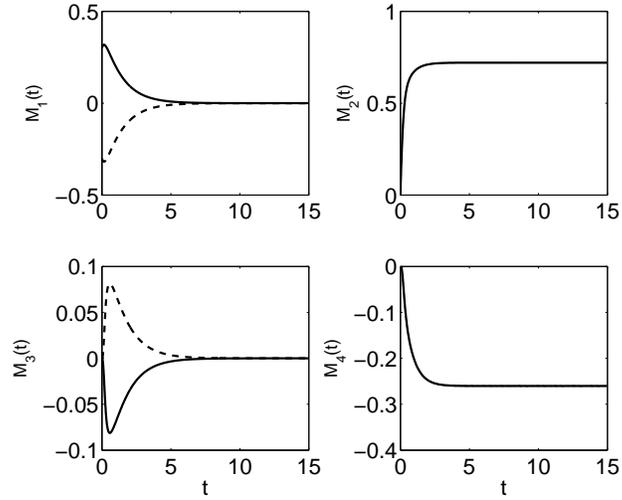}
    \caption{Behavior of the first four diagonal cumulants obtained
     from the truncated set of eleven equations for the cumulants for $N=5$, $D=1.33$,
     $\theta=2$ and two different initial conditions: $M_1(0)=0.3$ (solid lines); $M_1(0)=-0.3$ (dashed lines) } 
\label{fig:1}       
\end{figure}

\begin{figure}
\center        
\includegraphics[width=9cm]{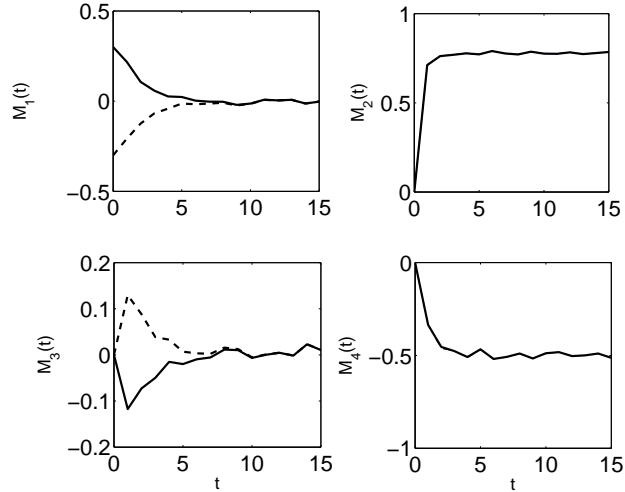}
\caption{Behavior of the first four diagonal cumulants obtained
     from the simulations of Langevin equations for the cumulants for $N=5$, $D=1.33$,
     $\theta=2$ and two sets of initial conditions: $M_1(0)=0.3$ (solid lines); $M_1(0)=-0.3$ (dashed lines). $M_2(t)$ and $M_4(t)$ for the two sets of initial conditions are indistinguishable with the graph resolution.}
\label{fig:2} 
\end{figure}

Let us first investigate what happens for parameter values such that, in the infinite size limit,  Desai and Zwanzig obtained a single stable stationary first moment.
In Fig.\ \ref{fig:1} we depict the 
behavior of the first four diagonal cumulants obtained from the truncated hierarchy of eleven equations for the cumulants for $N=5$, $D=1.33$, $\theta=2$ and two sets of initial conditions: $M_1(0)=0.3$ (solid lines) and $M_1(0)=-0.3$ (dashed lines). From the numerical simulations of Langevin equations for $N=5$, $D=1.33$, $\theta=2$ and the same two sets of initial conditions we get the results depicted in Fig.\ \ref{fig:2}. The long-time limit results are independent of the initial preparation and, except for the long time value of $M_4$, the truncated hierarchy yields a time evolution of the diagonal moments very much in agreement with those obtained with the Langevin simulations. We have numerically analyzed other set of parameter values and initial conditions  and the numerical findings lead us to conclude that,
for parameter values such that $M_1^{st}=0$ is the only stable stationary solution,  the set of eleven equations  represents a reliable approximation to the correct behavior for systems of even very modest sizes.  

\begin{figure} 
\center
\includegraphics[width=9cm]{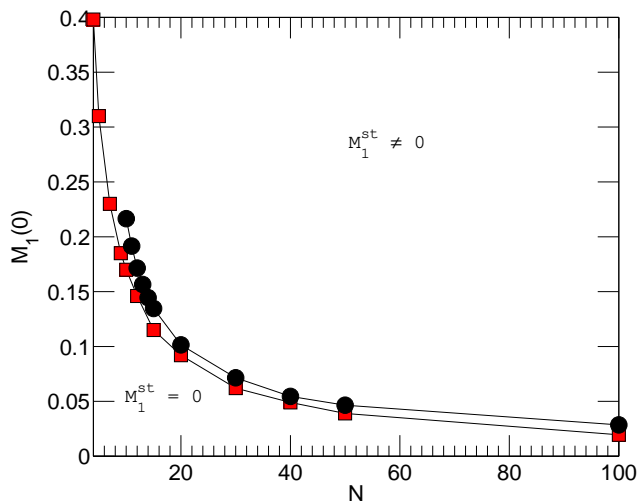} 
   \caption{(Color online) The (black) dots and the (red) squares represent the limiting initial values of $M_1(0)$ vs. $N$ separating the region of stability of the long time zero solution of the hierarchy of equations for the cumulants truncated at the fourth order or the second order levels respectively. Note that for $N<10$, $M_1^{st}=0$ for the fourth order truncation regardless of the initial preparation, while for the second order truncation, this fact is true for $N<4$. The initial conditions are $M_1(0) \neq 0$ and all the other cumulant are taken to be zero initially. The lines are a guide to the eye. System parameters: $D=0.33$ and $\theta=2$.  }
\label{fig:ciN}
\end{figure}

Let us next consider a set of values for the parameters $\theta$ and $D$ such that, for infinite systems, Desai and Zwanzig \cite{DZ1978} found that truncation of the hierarchy of cumulant moments leads, in the $N\rightarrow \infty $ limit, to two coexisting stable stationary nonzero values for the first moment while $M_1^{st}=0$ is unstable. In the case of small finite systems,  the numerical solution of the truncated hierarchy still leads to three stationary values but $M_1^{st}=0$ might be stable for a range of initial conditions. This range depends on the system size, the parameter values and the level of truncation. In Fig. \ref{fig:ciN}, we depict the results for $M_1^{st}$
obtained from the long time solution of the hierarchy of cumulant equations truncated at the fourth order level (black dots) and at the second order level (red squares) with $M_1(0) \neq 0$ and all the other cumulant set to zero. For all values of $N$ considered, we have used $D=0.33$ and $\theta=2$. With the fourth order truncation, the long time solution $M_1^{st}=0$ is stable for $N < 10$ regardless of the initial condition. For $N> 10$, the stability of the zero solution depends on the initial value $M_1(0)$. 
With the second order truncation, the solution $M_1^{st}=0$ is stable for $N < 4$, while it becomes unstable for some range of initial conditions as $N>4$. Thus, there is an influence of the level of truncation in the stability diagram of the zero stationary solution.

\begin{figure}
\center            
\includegraphics[width=9cm]{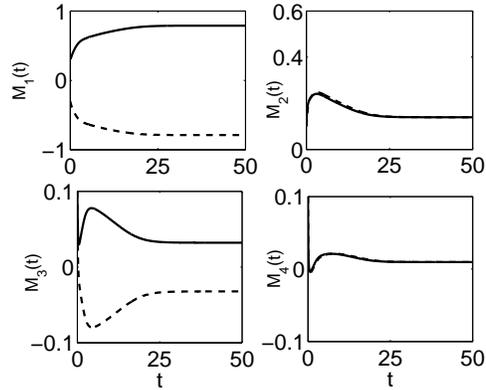}
\caption{Behavior of the first four diagonal cumulants obtained
     from the truncated hierarchy of equations for the cumulants for $N=15$, $D=0.33$,
     $\theta=2$ and two sets of initial conditions, $M_1(0)=0.3$ (solid lines); $M_1(0)=-0.3$ (dashed lines).}
\label{cuN15}
\end{figure}

In Fig.\ \ref{cuN15} we depict the results for the time evolution of the
first four diagonal cumulants obtained from the numerical solution of
the truncated hierarchy of equations for $N=15$,
$D=0.33$, $\theta=2$ for two sets of initial conditions ($M_1(0)=\pm 0.3$).

\begin{figure}
\center
\includegraphics[width=9cm]{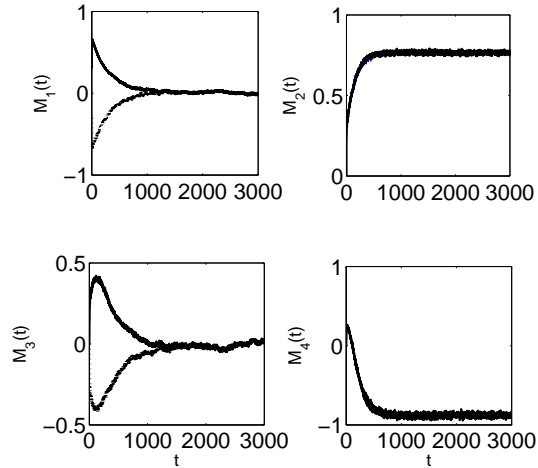}
\caption{Behavior of the first four diagonal cumulants obtained
     from the simulations of Langevin equations for the cumulants for $N=15$, $D=0.33$,
     $\theta=2$ and two sets of initial conditions.}
\label{siN15}
\end{figure}

In Fig.\ \ref{siN15} we depict the behavior of the first four diagonal
cumulants as obtained from numerical simulations of the Langevin
equations for a system with $N=15$, $D=0.33$ and $\theta = 2$ with the
two sets of initial conditions. The truncated set of cumulant equations leads to two nonzero first moment stationary values. The discrepancies  between the results in  Figs.\ \ref{cuN15} and \ref{siN15} for the same parameter values are evident. The Langevin simulation results indicate that
the long time behavior of the cumulants is independent of the initial
condition. This fact is consistent with an exact result: according to the
H-theorem \cite{RISKEN1984}, the long time equilibrium solution of
Eq.\ (\ref{EQ3}) is independent of the initial preparation of the
system. The first moment, $M_1(t)$, has therefore a single stationary
value. Considering the canonical form of the long time solution of
Eq.\ (\ref{EQ3}) and the symmetry of the potential in
Eq.\ (\ref{EQ4}) we have that, for any finite system, $M_1^{st}=0$, which is the long time value obtained via Langevin simulations.  Then, the stability of the nonzero long time solutions and the dependence of the initial preparation seem to be an artifact
of the truncation rather than a property of a finite system. 

\begin{figure} 
\center
\includegraphics[width=9cm]{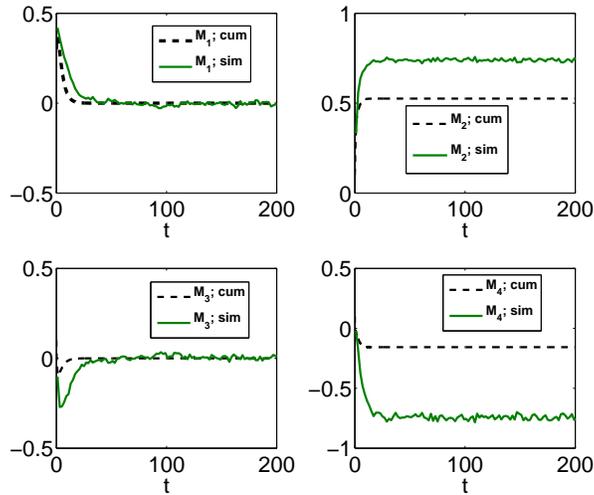}
\caption{(Color online) Behavior of the first four diagonal cumulants estimated
from the numerical simulations of Eq.\ (\ref{EQ1}) for $N=3$ (solid green lines), as well as from the truncated hierarchy of equations for the
cumulant moments (black dashed lines). Other parameter values
$D=0.33$ and $\theta=2$.}
\label{fig:cumsimN3}
\end{figure}

Even for $\theta$ and $D$ values such that multiple steady solutions are possible, the solution $M_1^{st}=0$ of the truncated hierarchy is the only stable stationary solution for $N$ sufficiently small ($N<10$ for the parameter values in Fig.\ \ref{fig:ciN}). It is worth to compare for these small systems the time evolution of the first few cumulants obtained with the truncated hierarchy with the one provided by Langevin simulations.  In Fig.\ \ref{fig:cumsimN3} we depict the results of the evolution of the first four diagonal cumulants obtained with the numerical simulations of Eq.\ (\ref{EQ1}) for $N=3$, as well as 
 from the truncated hierarchy of equations for $\theta=2$,  $D=0.33$. We see that, as expected, for this small system the truncated hierarchy leads to $M_1^{st}=0$, in agreement with the results obtained from the Langevin simulations. In contrast, the second and fourth order cumulants steady values obtained from the truncated hierarchy and from the Langevin simulations show large quantitative differences, indicating the limitations of the truncated hierarchy. 

\begin{figure} 
\center
\includegraphics[width=9cm]{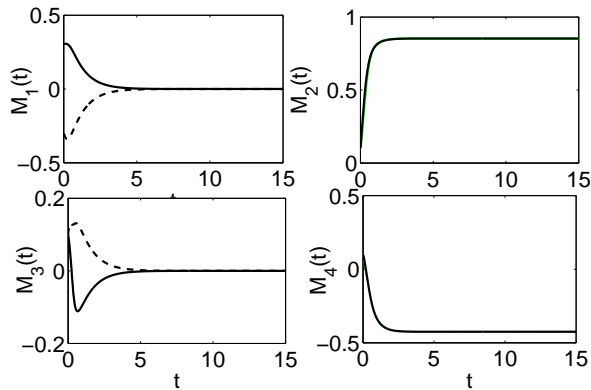} 
\caption{Behavior of the first four diagonal cumulants of the
     truncated hierarchy of equations for $N=10$,
     $D=0.5$, $\theta=-0.5$ for two sets of initial conditions.  }
\label{fig:cum11Qmp5}
\end{figure}

\begin{figure} 
\center
\includegraphics[width=9cm]{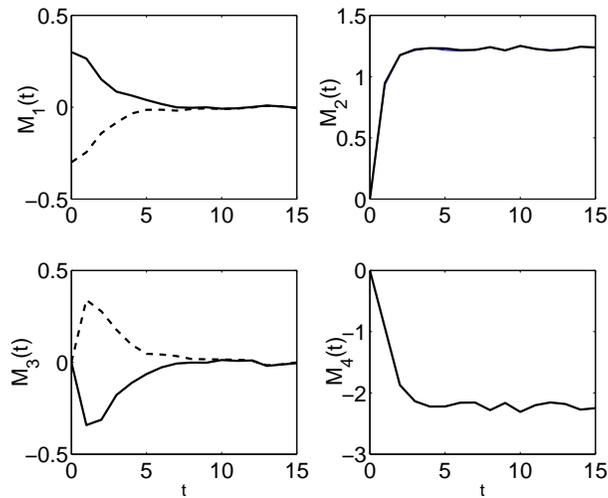} 
\caption{(Color online) Behavior of the first four diagonal cumulants obtained
     from numerical simulations of the Langevin equations for $N=10$,
     $D=0.5$, $\theta=-0.5$ for two sets of initial conditions.  }
\label{fig:cumsimQmp5}
\end{figure}

We now turn our attention to the model with a negative global
coupling parameter $\theta < 0$. Figs. \ref{fig:cum11Qmp5} and
\ref{fig:cumsimQmp5} show the results obtained, respectively, with the
truncated hierarchy of equations for the cumulant moments, and with
the full simulation of the Langevin equations for systems with $N=10$,
$D=0.5$ and $\theta=-0.5$. We see that the long time limit is
independent of the initial conditions. There are differences in the
long time value for the second and fourth cumulant moments obtained
with the hierarchy of equations and the simulations. The relaxation
towards the stationary solution is quite fast, although the relaxation
time with the full simulation is somewhat longer.  Even though the
results in Fig.\ \ref{fig:cum11Qmp5} differ quantitatively from the
simulations results in Fig.\ \ref{fig:cumsimQmp5}, they yield a good
qualitative approximation.
An extensive numerical analysis of the truncated
hierarchy indicates that for $\theta<0$ there is only a single zero stationary value
for the first moment regardless the value of $N$. 

\section{Conclusion}
\label{conclu}
In conclusion, the work presented here indicates that care must be taken when using an approximation to the dynamical behavior in a chain of interacting identical objects, based on truncating 
the infinite hierarchy of cumulant moments. Even for very large systems, if the parameter values considered are such that the truncated hierarchy leads to two stable coexisting solutions, the approximation is not correct. The results of the numerical simulations of the Langevin equations and the exact properties of the Fokker-Planck equation for finite systems of any size indicate that the coexistence of two stationary solutions is an artifact of the truncation. On the other hand, when the truncated hierarchy has a single stationary stable solution, it provides a reliable approximation of the system dynamics even for systems of very modest size. Although our conclusions are based on the study of a particular model, we think that they are qualitatively relevant for other cases. Langevin dynamics with additive noise are good
general representations of dynamical systems in contact with a thermal
environment. Actually, Langevin dynamics are more realistic than their
deterministic limits often used to study dissipative nonlinear
systems, where dissipation is included but all fluctuations are
neglected. Global interactions are also quite general. The quartic nonlinearity we
study here is a particular case. But having a different nonlinearity should
not change our qualitative conclusions. The nonlinearity considered
includes already the nontrivial effects associated to the dynamical
coupling between noise and nonlinearity.


\begin{thebibliography}{99.}

\bibitem{RISKEN1984}
H. Risken, \textit{The Fokker-Planck Equation},
(Springer, Berlin Heidelberg New York 1984).

\bibitem{mar} J. Marcinkiewicz, Math. Z. \textbf{19}, 612 (1938).

\bibitem{hantal} P. H\"{a}nggi and P. Talkner,  J. Stat. Phys. {\bf 22}, 65 (1980).



\bibitem{DZ1978}
R.C. Desai, R. Zwanzig, J. Stat. Phys. {\bf 19}, 1 (1978).



\bibitem{KomShi1975} 
K. Kometani and H. Shimizu, J. Stat. Phys.{\bf 13}, 473 (1975).

\bibitem{CDGMH2003} 
J. Casado-Pascual, C. Denk, J. G\'omez-Ord\'o\~nez, M. Morillo, P.H\"{a}nggi, Phys. Rev. E {\bf 67}, 036109 (2003).


\end{thebibliography}
\end{document}